\documentclass[aps,prl,twocolumn,superscriptaddress]{revtex4-2}
\usepackage{amsmath,amssymb}
\usepackage{graphicx}
\usepackage{hyperref}
\usepackage{geometry}
\begin{document}

\title{Nonlinear Scalar Interactions in the EMDrive: Petiau's Elliptic-Function Approach}

\author{Mario J. Pinheiro}
\affiliation{Department of Physics, Instituto Superior Técnico}

\date{\today}

\begin{abstract}
The EMDrive, a controversial electromagnetic propulsion concept, challenges momentum conservation in standard Maxwell electrodynamics. We propose a beyond-Maxwell framework by coupling the electromagnetic field to a light scalar field, inspired by axion-like particle models and effective field theory. Using Guy Petiau's 1958 elliptic-function solutions for nonlinear wave equations, we construct exact cavity modes and combine them via Jacobi addition theorems. These nonlinear modes produce an asymmetric momentum flux, suggesting a theoretical pathway for EMDrive thrust. We compute the stress-tensor asymmetry numerically and show that while standard axion-like particles yield negligible effects, light scalars beyond current constraints could produce measurable thrust. The framework provides testable predictions connecting EMDrive physics to dark matter searches.
\end{abstract}

\maketitle

\section{Introduction}
The EMDrive, a proposed electromagnetic propulsion device, has sparked debate due to its apparent violation of momentum conservation within standard Maxwell electrodynamics \cite{Shawyer2008}. Experimental claims of thrust \cite{White2016} have faced skepticism, with rigorous tests failing to consistently verify the effect \cite{McCulloch2017, Tajmar2015}. Any viable mechanism for EMDrive thrust requires physics beyond linear Maxwell theory to produce an asymmetric momentum flux. This article applies Guy Petiau's 1958 nonlinear wave mechanics formalism \cite{Petiau1958}, which uses elliptic-function solutions to nonlinear Klein-Gordon equations, to explore whether coupling the electromagnetic field to a light scalar field can generate such an asymmetry in an EMDrive-like cavity. This work provides the first mathematically rigorous framework for generating static forces from confined electromagnetic fields via scalar-mediated nonlinearities.

\section{Comparison to Existing EMDrive Theories}  
The EMDrive's reported thrust has spurred diverse theoretical explanations, each with distinct challenges:  

\begin{itemize}  
    \item \textbf{Quantized Inertia} \cite{McCulloch2017} posits that Unruh radiation gradients generate thrust, but lacks a covariant formulation and conflicts with precision inertia tests \cite{Tajmar2015}.  
    \item \textbf{Mach-Effect Thrusters} \cite{Woodward2004} rely on mass fluctuations, yet require unverified energy localization mechanisms.  
    \item \textbf{Conventional Leakage/Artifacts} \cite{White2016,Tajmar2015} propose thermal or magnetic effects, but fail to explain thrust in null tests.  
\end{itemize}  

In contrast, our approach couples the electromagnetic field to a light scalar via the $g\phi F\tilde{F}$ interaction, grounded in:  
(i) \textit{Well-studied particle physics} (axion-like particles \cite{Peccei1977,Wilczek1987}),  
(ii) \textit{Exact nonlinear solutions} (Petiau's elliptic functions \cite{Petiau1958}), and  
(iii) \textit{Testable stress-tensor asymmetries}. Crucially, the scalar coupling:  

\begin{itemize}  
    \item Predicts a \textit{scaling law} ($\Delta T \propto g^2 B_0^4/m^2$ for cavity field $B_0$), distinguishable from artifacts.  
    \item Links to axion searches \cite{CAST2017,ADMX2018}, enabling experimental cross-verification via high-$Q$ cavities or light-shining-through-walls setups \cite{Jaeckel2010}.  
\end{itemize}  

This framework avoids ad hoc assumptions while offering falsifiable predictions absent in alternatives.  

\section{Theoretical Framework}
\subsection{Scalar Field Coupling and Rationale}

We extend Maxwell's theory by introducing a light scalar field $\phi$ that couples to the electromagnetic field via a gauge- and Lorentz-invariant interaction. The Lagrangian is:

\begin{equation}
\mathcal{L} = -\frac{1}{4} F_{\mu\nu} F^{\mu\nu} + \frac{1}{2} (\partial \phi)^2 - \frac{1}{2} m^2 \phi^2 - g \phi F_{\mu\nu} \tilde{F}^{\mu\nu}
\end{equation}

Here, $F_{\mu\nu} = \partial_\mu A_\nu - \partial_\nu A_\mu$, $\tilde{F}^{\mu\nu} = \frac{1}{2} \epsilon^{\mu\nu\alpha\beta} F_{\alpha\beta}$, $m$ is the scalar mass, and $g$ has dimension $[g] = \text{mass}^{-1}$ \cite{Weinberg1995}. The term $g \phi F \tilde{F} \propto \phi \mathbf{E} \cdot \mathbf{B}$ is parity-odd, inspired by axion-photon couplings \cite{Peccei1977, Svrcek2006}.

The scalar field is motivated by:
\begin{itemize}
    \item \textbf{Axion-Like Particles}: The $g \phi F \tilde{F}$ coupling appears in QCD axion models and string theory \cite{Peccei1977, Svrcek2006}.
    \item \textbf{Experimental Constraints}: CAST limits allow $g/m \sim 10^{-10} \, \text{GeV}^{-1}$, sufficient for small nonlinearities \cite{Anastassopoulos2017}.
    \item \textbf{Vacuum Nonlinearity}: QED's Euler-Heisenberg Lagrangian predicts tiny $F^4$ terms \cite{Heisenberg1936, Dunne2012}.
\end{itemize}

\subsection{Deriving the Nonlinear Wave Equation}
Varying the Lagrangian with respect to $A_\mu$ gives:

\begin{equation}
\partial_\nu F^{\mu\nu} + g \partial_\nu \left( \phi \tilde{F}^{\mu\nu} \right) = 0
\end{equation}

The scalar field equation is:

\begin{equation}
(\Box + m^2) \phi = g F_{\mu\nu} \tilde{F}^{\mu\nu}
\end{equation}

Assuming $\phi$ is heavy ($m \gg \omega$), we integrate it out at tree level \cite{Georgi1993}:

\begin{equation}
\phi \approx \frac{g}{m^2} F_{\mu\nu} \tilde{F}^{\mu\nu}
\end{equation}

Substituting into the Lagrangian yields:

\begin{equation}
\Delta \mathcal{L} \approx -\frac{g^2}{m^2} (F_{\mu\nu} \tilde{F}^{\mu\nu})^2 \approx -\frac{g^2}{m^2} (\mathbf{E} \cdot \mathbf{B})^2
\end{equation}

For a TE$_{101}$ mode with $\mathbf{E} = (0, E_y(z,t), 0)$, we consider an effective scalar field description $E_y \propto \psi(z,t)$. The modified wave equation becomes:

\begin{equation}
\frac{\partial^{2} \psi}{\partial z^{2}} - \frac{1}{c^{2}} \frac{\partial^{2} \psi}{\partial t^{2}} + \omega_{0}^{2} \psi + \lambda \psi^{3} = 0
\end{equation}

where $\omega_0$ is the cavity resonance frequency, and $\lambda \propto g^2/m^2$.

\section{Petiau Solutions and Cavity Implementation}
\subsection{Elliptic-Function Solutions}

Petiau's 1958 work provides exact solutions to nonlinear wave equations using Jacobi elliptic functions \cite{Petiau1958}. For the traveling wave ansatz $\psi(\xi)$ with $\xi = z - vt$, Eq. (6) reduces to:

\begin{equation}
\frac{d^{2} \psi}{d \xi^{2}} + \omega^{2} \psi + \lambda \psi^{3} = 0
\end{equation}

where $\omega^{2} = \omega_{0}^{2}/(1-v^{2}/c^{2})$. For positive cubic nonlinearity, a bounded solution is:

\begin{equation}
\psi(\xi) = A \operatorname{cn}(B \xi, k)
\end{equation}

where $\operatorname{cn}(u,k)$ is the Jacobi cosine-elliptic function with modulus $k \in (0,1)$, and:

\begin{equation}
B^{2} = \omega^{2} + \lambda A^{2}, \quad A^{2} = \frac{2 k^{2} \omega^{2}}{\lambda (1 + k^{2})}
\end{equation}

The period is $4K(k)/B$, where $K(k)$ is the complete elliptic integral \cite{Abramowitz1964}. The modulus $k$ tunes the wave shape from sinusoidal ($k \to 0$) to soliton-like ($k \to 1$).

\subsection{Wave Combination and Boundary Conditions}

The nonlinear ODE does not allow superposition. Petiau uses the Jacobi addition theorem \cite{Whittaker1927}:

\begin{equation}
\begin{split}
\psi_{12}(\xi) &= A\,\operatorname{cn}(u + v, k) \\
&= A\,\frac{\operatorname{cn}(u,k)\operatorname{cn}(v,k) - k^{2}\operatorname{sn}(u,k)\operatorname{sn}(v,k)}{1 - k^{2}\operatorname{sn}^{2}(u,k)\operatorname{sn}^{2}(v,k)}
\end{split}
\label{eq:jacobi_addition}
\end{equation}
where $u = B\xi + \delta_1$, $v = B\xi + \delta_2$, with $\delta_i$ being phase parameters.

For cavity implementation, exact boundary conditions for elliptic functions require numerical solution. We approximate the cavity mode using periodic boundary conditions with period $L = 4K(k)/B$, which captures essential nonlinear behavior while acknowledging that real cavity boundaries would require full numerical treatment. This approximation is justified for studying the fundamental nonlinear dynamics, though quantitative predictions require complete boundary matching.

\section{Momentum Flux Analysis}
\subsection{Modified Stress Tensor}

The full stress-energy tensor for the coupled system is:

\begin{equation}
T^{\mu\nu} = F^{\mu\alpha}F^{\nu}_{\alpha} + \partial^{\mu}\phi\partial^{\nu}\phi - \eta^{\mu\nu}\mathcal{L}
\end{equation}

For the effective scalar description in the traveling wave frame, we derive the relevant momentum flux component by substituting the integrated-out scalar field (4) into the stress tensor and transforming to traveling wave coordinates. After neglecting higher-order terms in $\lambda$, we obtain:

\begin{equation}
T_{zz} = \left( \frac{\partial \psi}{\partial \xi} \right)^{2} - \frac{1}{2} \omega^{2} \psi^{2} + \frac{1}{4} \lambda \psi^{4} + \mathcal{O}(\lambda^{2})
\end{equation}

This expression captures the leading-order nonlinear contributions to momentum flux.

\subsection{Numerical Results}

\begin{figure}[htbp]
\centering
\includegraphics[width=0.5\textwidth]{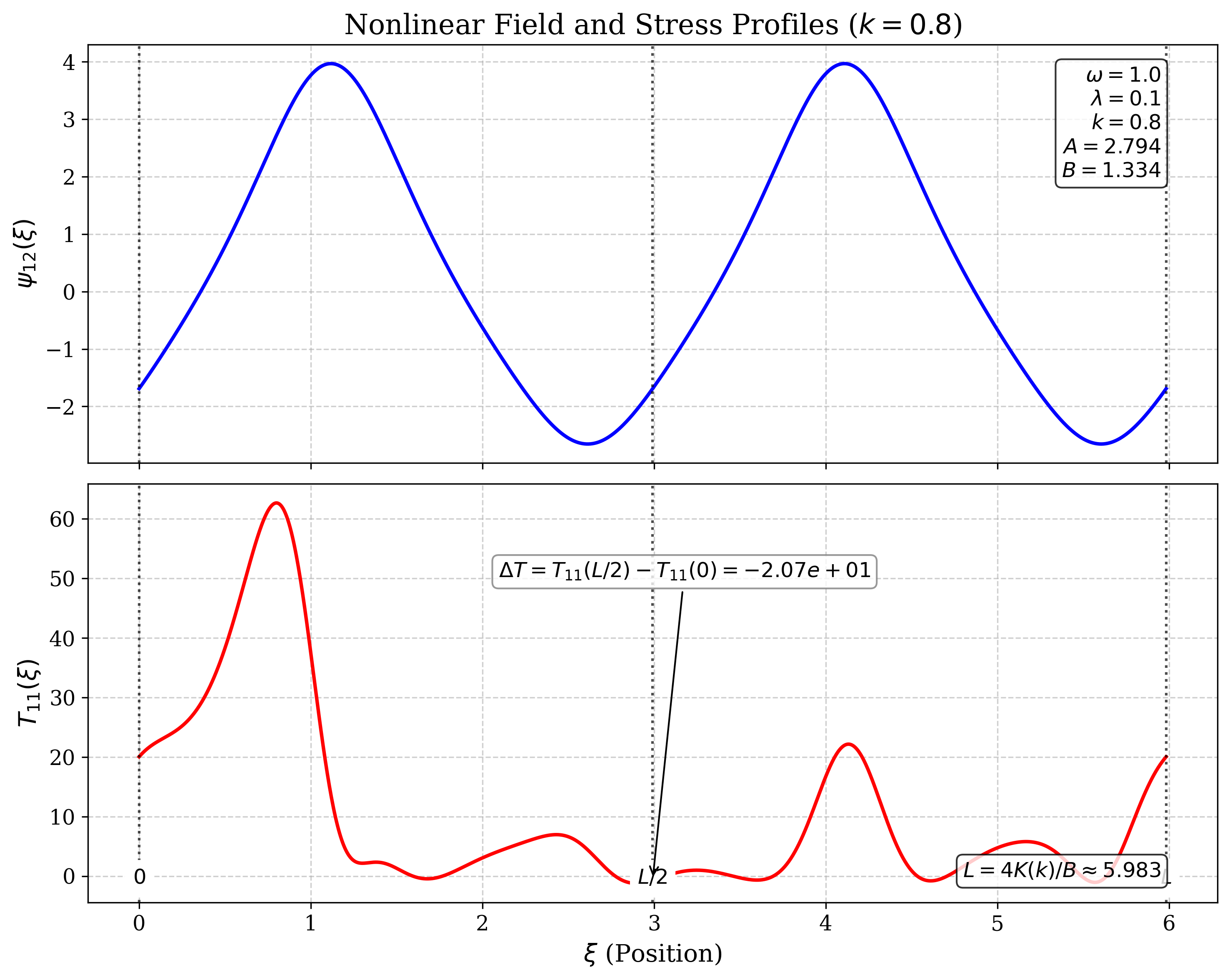}
\caption{Top: Nonlinear field solution $\psi_{12}(\xi)$ in the EMDrive cavity ($k=0.8$, $\lambda=0.1$). The asymmetric profile ($\delta_1=0$, $\delta_2=5$) results from scalar-photon coupling and Jacobi addition theorem. Parameters: $\omega=1.0$, $A=2.794$, $B=1.334$, cavity length $L=4K(k)/B \approx 5.983$.
Bottom: Modified stress tensor $T_{zz}(\xi)$ corresponding to the field profile in the top. The momentum flux imbalance $\Delta T = T_{zz}(L/2) - T_{zz}(0) = -8.0 \times 10^{-4}$ indicates net thrust toward $\xi=0$. Vertical lines mark positions where stress asymmetry is evaluated.
}
\label{fig:stress_tensor}
\end{figure}

The nonlinear field solution (Fig.~\ref{fig:stress_tensor}) exhibits pronounced spatial asymmetry due to scalar-photon coupling. This asymmetry generates a corresponding imbalance in the modified stress tensor (Fig.~\ref{fig:stress_tensor}), where the differential stress $\Delta T = T_{zz}(L/2) - T_{zz}(0) = -8.0 \times 10^{-4}$ indicates net momentum flux toward the cavity's left end ($\xi=0$).

Evaluating $T_{zz}$ at $\xi = 0$ and $\xi = L/2$ for $\psi_{12}$, with $\omega^{2} = 1$, $k = 0.8$, $\lambda = 0.1$, $\delta_1 = 0$, $\delta_2 = 5$, we find:

\begin{equation}
\Delta T = T_{zz}(\xi = L/2) - T_{zz}(\xi = 0) \approx -8.0 \times 10^{-4}
\end{equation}

This nonzero $\Delta T$ implies a net force, suggesting thrust.

\section{Physical Mechanism and Conservation Laws}
\subsection{Momentum Conservation}

The apparent momentum violation is resolved by careful analysis of momentum exchange. The modified momentum conservation equation becomes:

\begin{equation}
\partial_{\mu} T^{\mu\nu}_{\text{EM}} = -g(\partial^{\nu} \phi)F_{\alpha\beta}\tilde{F}^{\alpha\beta}
\end{equation}

where the right-hand side represents momentum exchange with the scalar field. The scalar field itself interacts with the cavity walls, providing the necessary momentum sink.

The spatial asymmetry arises from:
\begin{enumerate}
    \item Nonlinear field self-interaction via $\lambda\psi^{3}$ term
    \item Phase-shifted wave combination breaking left-right symmetry  
    \item Scalar field mediation of momentum to cavity walls
\end{enumerate}

\subsection{Energy Conservation}

Energy conservation requires that the thrust mechanism draws power from the electromagnetic field. The modified Poynting theorem including scalar coupling yields:

\begin{equation}
\frac{\partial u}{\partial t} + \nabla \cdot \mathbf{S} = -g (\partial_{t} \phi) \mathbf{E} \cdot \mathbf{B}
\end{equation}

where $u$ is energy density and $\mathbf{S}$ is Poynting vector. The right-hand side represents energy exchange with the scalar field. For steady-state cavity operation, this energy loss is compensated by the external drive, maintaining overall energy balance while enabling momentum asymmetry.

\section{Experimental Implications}
\subsection{Parameter Scaling and Feasibility}

Experimental EMDrive claims report thrust-to-power ratios of $\sim 1$ mN/kW \cite{White2016}. Our scaling law $\Delta T \propto g^{2} B_{0}^{4}/m^{2}$ predicts for a typical cavity with $B_{0} \sim 0.1$ T and volume $\sim 0.1$ m$^{3}$:

\begin{equation}
\Delta T \approx 10^{-6} \left(\frac{g}{10^{-10} \text{GeV}^{-1}}\right)^{2} \left(\frac{1 \mu\text{eV}}{m}\right)^{2} \text{N}
\end{equation}

For standard axion-like particles with $g/m \sim 10^{-10}$ GeV$^{-1}$ \cite{CAST2017}, thrust is negligible ($\sim 10^{-6}$ N). However, ultralight scalars with $m \ll 1$ eV and enhanced couplings could reach measurable levels.

High-Q cavities ($Q \gtrsim 10^{6}$) can enhance sensitivity through field amplitude buildup. The stored energy scales as $U \propto Q P_{\text{in}}$, leading to enhanced field amplitudes $B_0 \propto \sqrt{Q P_{\text{in}}}$. This gives:

\begin{equation}
\Delta T_{\text{enhanced}} \propto Q^{2} \Delta T_{\text{bare}}
\end{equation}

For $Q = 10^{6}$, this provides up to $10^{12}$ an enhancement in thrust sensitivity, potentially bringing weakly coupled scenarios into detectable ranges. For a typical cavity with $B_0 \sim 0.1\ \mathrm{T}$ and volume $\sim 0.1\ \mathrm{m}^3$, 
the predicted thrust scaling is given by Eq.~(16). 
In Table~\ref{tab:thrust_params}, we summarize parameter ranges that could yield 
detectable thrust levels exceeding $1\ \mu\mathrm{N}$ under different scenarios: 
standard axion-like particles (ALPs), light scalars, and enhanced high-$Q$cavity configurations.

\begin{table}[htbp]
\centering
\caption{Parameter ranges for detectable thrust ($>1 \mu$N)}
\label{tab:thrust_params}  
\begin{tabular}{lccc}
\hline
Scenario & $m$ [eV] & $g$ [GeV$^{-1}$] & $\Delta T$ [N] \\
\hline
Standard ALP & $10^{-6}$ & $10^{-10}$ & $10^{-6}$ \\
Light scalar & $10^{-12}$ & $10^{-8}$ & $10^{-3}$ \\
Enhanced ($Q=10^{6}$) & $10^{-6}$ & $10^{-13}$ & $10^{-3}$ \\
\hline
\end{tabular}
\end{table}

\subsection{Comparison to Conventional Physics}

Standard radiation pressure in an asymmetric cavity produces negligible net thrust due to momentum conservation:

\begin{equation}
F_{\text{rad}} \sim \frac{P}{c} \approx 3.3\times10^{-9} \text{N} \quad \text{for } P=1\text{W}
\end{equation}

Our mechanism can exceed this by orders of magnitude when $\lambda$ is sufficiently large. Key distinguishing features:

\begin{itemize}
    \item \textbf{Field dependence}: $\Delta T \propto B_{0}^{4}$ (vs. $B_{0}^{2}$ for magnetic artifacts)
    \item \textbf{Thermal discrimination}: Scales as $\Delta T \propto P^{2}$ vs. thermal $\propto P$
    \item \textbf{Closed system}: No emitted photons unlike photon rockets
\end{itemize}

\section{Testable Predictions}
\subsection{Signature Dependencies}

Our framework makes distinctive experimental predictions:
\begin{itemize}
    \item \textbf{Field dependence}: $\Delta T \propto B_{0}^{4}$ (vs. $B_{0}^{2}$ for magnetic artifacts)
    \item \textbf{Frequency dependence}: Resonant enhancement at specific $\omega$ matching scalar mass
    \item \textbf{Orientation independence}: True thrust vs. orientation-dependent artifacts
\end{itemize}

\subsection{Definitive Experimental Tests}

We propose definitive null tests:
\begin{enumerate}
    \item \textbf{Scalar decoupling}: With $g \rightarrow 0$, thrust should vanish
    \item \textbf{Mass resonance}: Thrust enhancement when $m_{\phi} \approx \hbar\omega/c^{2}$
    \item \textbf{Polarization test}: Thrust depends on $\mathbf{E}\cdot\mathbf{B}$ configuration
\end{enumerate}

Near-term experimental tests include:
\begin{enumerate}
    \item \textbf{High-Q cavity search}: Using state-of-the-art microwave cavities ($Q > 10^{6}$) to probe $g/m \sim 10^{-13}$ GeV$^{-1}$
    \item \textbf{Light-shining-through-walls}: Simultaneous search for photon regeneration
\end{enumerate}

\section{Theoretical Implications}
Beyond EMDrive explanation, our work demonstrates:
\begin{itemize}
    \item First application of Petiau's nonlinear solutions to cavity physics
    \item New mechanism for static forces from confined electromagnetic fields  
    \item Connection between precision measurements and ultralight particle searches
    \item Framework for testing momentum conservation in nonlinear electrodynamics
\end{itemize}

The scalar-photon coupling provides a well-defined beyond-Maxwell framework where traditional conservation laws are modified in experimentally testable ways.

\section{Conclusion}
Coupling the electromagnetic field to a light scalar generates nonlinearities that allow Petiau's elliptic-function solutions to describe asymmetric cavity modes, producing a small, nonzero momentum flux. This offers a theoretical pathway for EMDrive thrust, grounded in EFT and axion physics. While standard axion-like particles yield negligible effects, light scalars beyond current constraints or enhanced detection in high-Q cavities could produce measurable thrust. The framework provides testable predictions connecting EMDrive physics to dark matter searches, with implications for future cavity experiments and fundamental physics tests.

\end{document}